# A QUANTITATIVE ANALYSIS OF WCAG 2.0 COMPLIANCE FOR SOME INDIAN WEB PORTALS

Manas Ranjan Patra[1], Amar Ranjan Dash[2], Prasanna Kumar Mishra[3]

[1]Department of Computer Science, Berhampur University, Berhampur 760 007, India
[2]Department of Computer Science, Berhampur University, Berhampur 760 007, India
[3]Department of Computer Science, Berhampur University, Berhampur 760 007, India

## ABSTRACT

*Web portals have served as an excellent medium to facilitate user centric services for organizations irrespective of the type, size, and domain of operation. The objective of these portals has been to deliver a plethora of services such as information dissemination, transactional services, and customer feedback. Therefore, the design of a web portal is crucial in order that it is accessible to a wide range of user community irrespective of age group, physical abilities, and level of literacy. In this paper, we have studied the compliance of WCAG 2.0 by three different categories of Indian web sites which are most frequently accessed by a large section of user community. We have provided a quantitative evaluation of different aspects of accessibility which we believe can pave the way for better design of web sites by taking care of the deficiencies inherent in the web portals.*

## 1. INTRODUCTION

In recent years, organizations have shown keen interest for their electronic presence by hosting web portals. Several web sites have been hosted in the web and its number is increasing in a phenomenal rate. Some web sites are really impressive while many are not so user-friendly. Thus, the design of a web site is crucial for it to be accessible by a wide range of user community. In the past several attempts have been made to develop standards to guide the design process of web sites and inclusion of certain basic features for its accessibility. The W3C has done pioneering work in this direction and has come out with a set of guidelines in the form of different versions of Web Component Accessibility Guidelines (WCAG), the most recent and standardized version being WCAG 2.0. Ideally, web sites should comply with each of the guidelines incorporated in WCAG 2.0. However, in practice it is found that many of the web portals do not adhere to the guidelines, thus causing difficulty in their usage.

Some studies have been carried out to evaluate the compliance of the guidelines provided in the earlier version WCAG 1.0 by different web sites. Abdulmohsen, Ali, Pam [1] evaluated the accessibility of the government web portals of Saudi Arabia and Oman, using WCAG 1.0 guidelines. The authors have manually checked the web portals and have verified the compliance of guidelines with respect to WCAG 1.0 and have used Multiweb, Lynx, and W3C Validation as software tools to find error percentages. They have used Assistive technologies and Haptic devices for users having problems like total blindness and mobility. But, they have not addressed

International Journal of Computer Science, Engineering and Applications (IJCSEA) Vol.4, No.1, February 2014

the issue of persons with color blindness. Irina and Ben [2] analyzed the home pages of 50 US state Web sites. The analysis was done manually and initial recommendations were made with 10 rules which did not include any recommendation for disable people.

In [3], Uthayas and Zahir presented a conceptual model to study e-government portals. The model is partitioned into two parts, first part consists of IS (Information system) evaluation and second part consist of WCAG 2.0 impact factors. The aim of the study was to see whether WCAG 2.0 is properly followed in e-government portals. Joanne, Dorothy, and Klaus [4] examined the disability levels in the form of accessibility from different continent that is European Union, Asia, and Africa. They use TAW as software tools to analyze accessibility status and WCAG 1.0 guidelines, to test check point levels. TAW describes a report of automatic and human issues [5][4]. In case of Jordan, they have evaluated Government websites by taking twenty five sites comprising of Ministry of Agriculture, Ministry of Water and Irrigation, Ministry of Transport. Approximately similar analysis was done by Malaysian researchers [6]. They have used WCAG 1.0 priority 1 and Bobby as automatic tool, to evaluate 9 websites. They have provided only the number of errors of different portals instead of computing the percentage of error. Thus, it doesn't give an exact picture of the overall error percentage which can be used as a comparative measure among different web portals.

Jeff and Mike present 4 categories of peoples having the problem of disabilities. Here they describe how different company developed software in order to increase accessibility of these people [7]. 15% to 20% of our population suffers from diseases like dyslexia which causes difficulty in accessing web sites. To help them access web portals without difficulty Vagner et al. [8] suggested 41 guidelines. According to the UN survey by 2050, 20% of the world's population will be above 60 years of age, many of them will be disabled. To overcome such age-related-functional limitations the authors categorized the problem into 4 types and analyzed the websites by using WAI AGE guideline and AARP heuristic [9]. Friederike and Dirk have evaluated websites using WAI methodology [10]. Chaomeng [11] described different strategies to improve the web design practices of Taiwan for disabled people, where they evaluated 35 national level web portals.

The rest of the paper is organized as follows: In section 2, we describe the features of WCAG 2.0 in detail which forms the basis for our study. Section 3 presents the methodology followed in our analysis. In section 4, we briefly describe three different categories of web portals used for our study. In section 5 we evaluate some web portals and in section 6 we present the results of our analysis. Section 7 provides the concluding remarks.

## 2. WCAG 2.0 GUIDELINES

WCAG (Web Component Accessibility Guideline) is an international standard for maintaining accessibility of different web portals. The first version of WCAG came out in the form of WCAG 1.0. Later, new features were added and the current version of WCAG 2.0 was recommended by W3C (World Wide Web Consortium) on 11 December 2008 which was approved by ISO as an international standard (ISO/IEC40500:2012) in October 2012.The comparative study\of WCAG 1.0 and WCAG 2.0 was done by Miquel, Mireia, Merce, Marc, Andreu, Pilar[12]. WCAG 2.0 comprises of a number of recommendations for making web contents more accessible to a wide range of people (including people with disabilities).

A schematic representation of the guidelines included in WCAG 2.0 is presented in figure 1 which in short describes the following:

- Web portals should provide text alternatives for all non text contents.
- Reading and navigation order should be logical and intuitive.
- Text and images of text should have a contrast ratio more than the value prescribed by WCAG 2.0.
- Web portals should provide image alternatives wherever text contents are hard to understand.
- All functionality of website should be available through keyboard.
- Web portals should not contain flashes/flickering that can cause seizures.
- While filling a form proper error messages should be prompted whenever there is an error

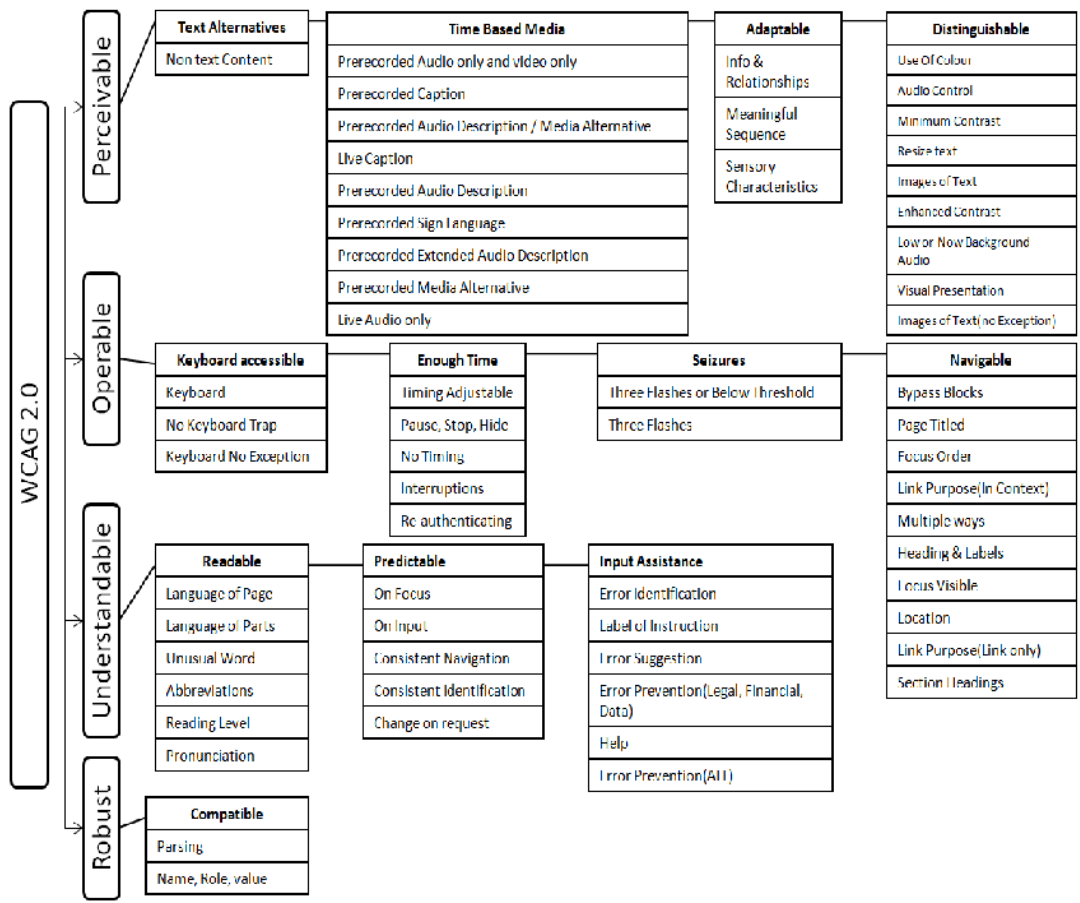

Figure 1: A Schematic representation of WCAG 2.0 Guidelines

Details of WCAG 2.0 guidelines can be found in [13]. Further, the guidelines are divided into three conformance levels, namely,

Level A: A webpage should satisfy a minimum set of requirements for its accessibility.
Level AA: A webpage should satisfy some additional features including those in Level A
Level AAA: A webpage should satisfy all the three levels of success Criteria, i.e., A, AA, and AAA)

## 3. METHODOLOGY

We have considered three different categories of websites for evaluating their accessibility. Under each category we have selected five web portals which are frequently used by the public, thus a total of 15 websites are used in our study. Next, the facilities provided by each of the websites are minutely checked. The testing parameters are based on WCAG 2.0 guidelines. The accessibility parameters are checked both manually and by using some tools.

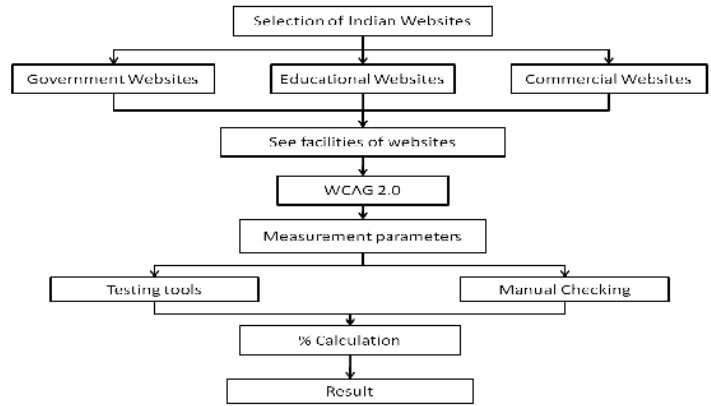

Figure 2: Web Accessibility Evaluation workflow

Table 1 shows some of the tools used in the measurement of different accessibility parameters in the websites. For checking the colour contrast of links we used "Check my colour". For checking the colour contrast of text and images of text we used two tools. In order to obtain the hexadecimal colour code we used "Colour Scheme studio" and then the colour contrast was checked with the help of the online tool "Colour Contrast Check". Different web browsers like Aurora, Comodo Dragon, and Google Chrome were used for accessing the web sites. We also used the "AChecker" tool for checking some of the parameters as per the WCAG 2.0 Guidelines.

| SL NO. | Target | Tools | URL |
|---|---|---|---|
| 1. | Colour Contrast (link) | Check my colour | http://www.checkmycolours.com/ |
| 2. | Colour Contrast (image's Text) | Colour Scheme studio | |
| | | Colour Contrast Check | http://snook.ca/technical/colour_contrast/colour.html |
| 3. | Web Browser | Aurora | |
| | | Chomodo Dragon | |
| | | Google Chrome | |
| 4. | Rule Wise Checker | AChecker | http://achecker.ca/checker/index.php |

Table 1: List of Tools

## 4. DESCRIPTION OF THE SAMPLE WEB SITES USED FOR ANALYSIS

This research is intended to provide an accessibility evaluation of 15 Indian websites based on WCAG 2.0. We choose 5 government websites. These government websites are associated with tax filling, taxation details, passport processing, Common Service Centre functions, and links to other Indian website (india.gov.in) etc. Similarly, we have chosen 5 educational websites, namely, two IITs, a central university, the University grant commission and a state level educational site. Educational websites facilitates online admissions, educational information dissemination, student registration, result publications etc. Finally, we have chosen 5 commercial websites, three of which are online shopping sites, one online railway ticketing site and an online trip planning and ticket booking site. The Commercial sites are frequently used for online shopping, online ticket booking, online transactions, and online product information etc. The data taken for our analysis is gathered during the period 15$^{th}$ Sep 2013 to 9$^{th}$ Oct 2013.

| Tested Websites | | |
|---|---|---|
| Government Websites | Educational Websites | Commercial websites |
| India.gov.in | www.iitb.ac.in | www.irctc.co.in |
| www.nic.in | www.iitd.ac.in | www.ebay.in |
| www.csc.gov.in | www.uohyd.ac.in | www.amazon.in |
| www.incometaxindia.gov.in | www.ugc.ac.in | www.flipkart.com |
| www.passportindia.gov.in | www.dheorissa.in | www.makemytrip.com |

Table 2: List of Websites

## 5. EVALUATION OF WEB PORTALS

### a. Government Sites

Figure 3.describes the overall average percentage violation per rule as provided in WCAG 2.0. In the following graph we analyze each accessibility issue for the Indian Government websites. The guideline where the % is 0 it means that either the web component which deals with that guideline is absent or there is no error for that web component.

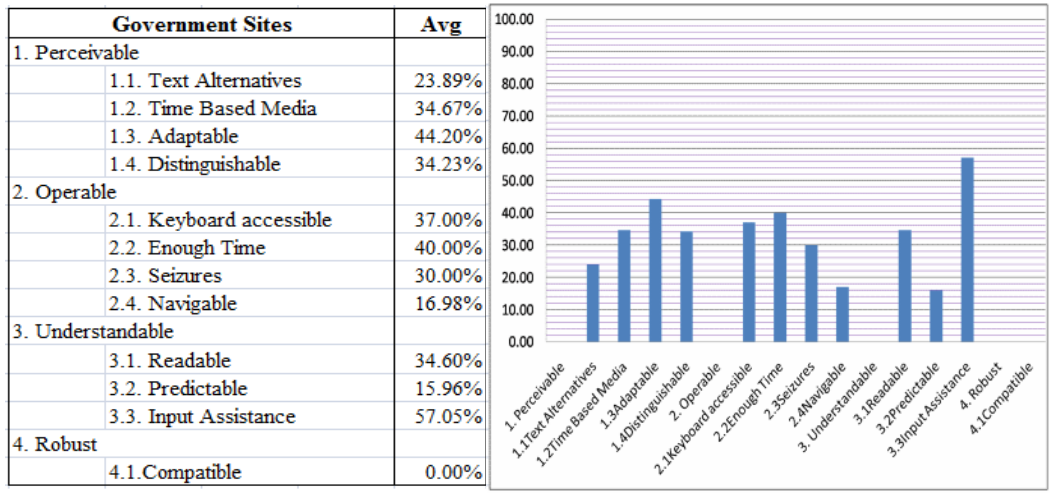

Figure 3: Violation of accessibility guideline in government websites

### i. Perceivable

While analyzing the perceivable features of the government websites, it is found that the maximum violated guideline is guideline 1.3 (Adaptable) in which government websites indicate 44.20% violation. Violations with respect to guideline 1.1 (Text Alternative), guideline 1.2 (Time Based Media), and guideline 1.4 (Distinguishable) were found to be 23.89%, 34.67%, and 34.23% respectively. Although the percentage is little less but this rule is violated by approximately all websites, so we provide a separate point by point graphical analysis in Figure 4. Here we find that the top 3 violated sub rules for distinguishability are guideline 1.4.6 (Enhanced contrast) with 68.25% violation, guideline 1.4.3 (Minimum contrast) with 65.75% violation, and guideline 1.4.8 (Visual Presentation) with 60.00% violation.

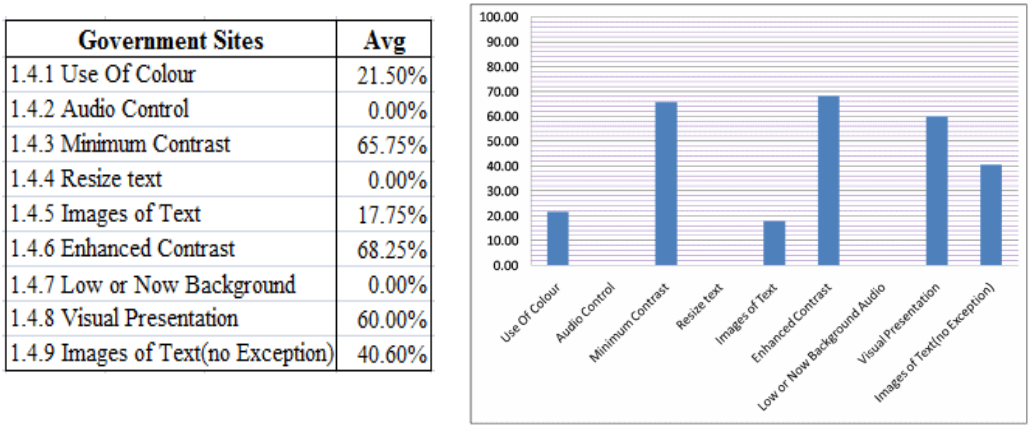

Figure 4: Violation of accessibility guideline (Distinguishable) in government websites

### ii. Operable

While evaluating for the operable principles, the maximum violated guideline is guideline 2.2 (Enough Time) in which government websites indicate 40.00% violation. But, for guideline 2.1 (Keyboard Accessible), guideline 2.3 (Seizures), and guideline 2.4 (Navigable) the violation percentages are 37.00%, 30.00%, and 16.98% respectively. Although the percentage is less but

International Journal of Computer Science, Engineering and Applications (IJCSEA) Vol.4, No.1, February 2014

this rule is violated by approximately all websites, so we give a separate point by point graphical analysis in Figure 5. Here we find that the top 3 violated sub rules for navigability are guideline 2.4.8 (Location) with 46.00% violation, guideline 2.4.10 (Section Headings) with 45.00% violation, and guideline 2.4.7 (Focus Visible) with 19.00% violation.

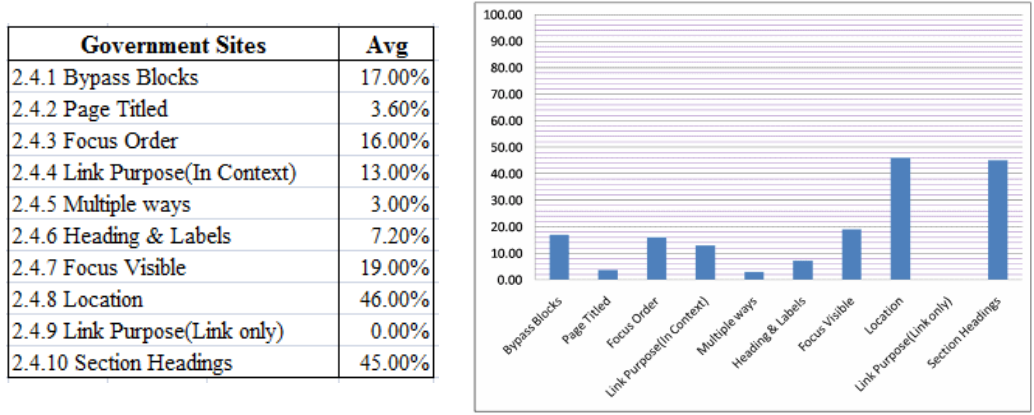

Figure 5: Violation of accessibility guideline (Navigable) in government website

### iii. Understandable

With respect to understandable principle in government websites, the maximum violated guideline is guideline 3.3 (Input Assistance) with 57.05% violation. Government websites indicate 34.60% and 15.96% violation in guideline 3.1 (Readable) and Guideline 3.2 (Predictable) respectively. This rule is violated by approximately all websites, so we give a separate point by point graphical analysis in Figure 6. Here we find that the top 2 violated sub rules of predictability are guideline 3.2.1 (On focus) with 32.00% violation and guideline 3.2.2 (On Input) with 26.00% violation.

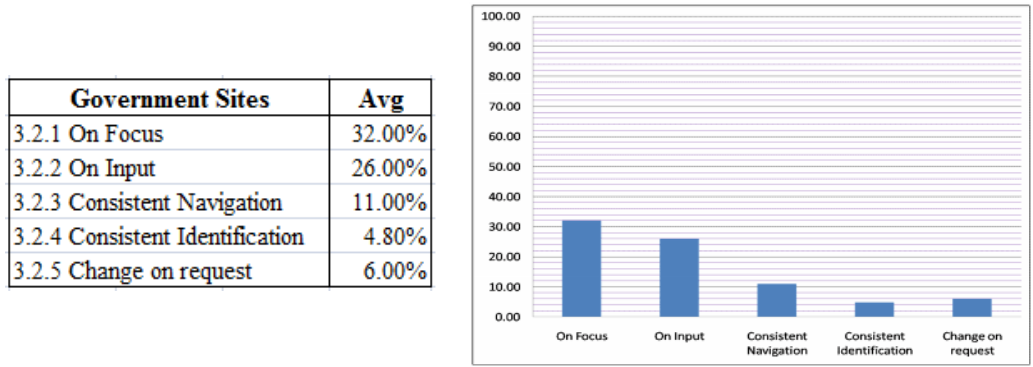

Figure 6: Violation of accessibility guideline (Predictable) in government website

### iv. Robust

According to our analysis the Indian Government Websites does not contain robust error.

### b. Educational Sites

Figure 7.describes the overall average percentage violation results per each rule of WCAG 2.0. In the following Graph we analyze each accessibility principle in Indian Educational websites.

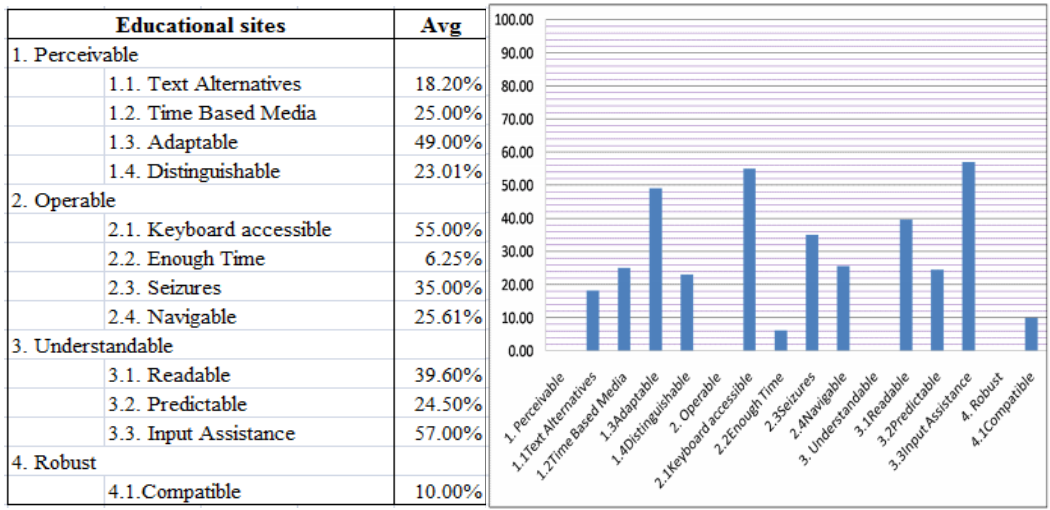

Figure 7: Violation of accessibility guideline in Educational websites

### i. Perceivable

In respect of perceivable principle in educational websites, the maximum violated guideline is guideline 1.3 (Adaptable) in which educational websites indicate 49.00% violation. Educational websites indicate 18.20%, 25.00%, and 23.01% violation in respect of guideline 1.1 (Text Alternative), guideline 1.2 (Time Based Media), and guideline 1.4 (Distinguishable) respectively. Although the percentage is small but this rule is violated by approximately all websites, so we introduce a separate point by point graphical analysis in Figure 8. Here we find that the top 3 violated sub rules for distinguishability are guideline 1.4.8 (Visual Presentation) with 48.00% violation, guideline 1.4.9 with 43.60% violation, and guideline 1.4.6 (Enhanced contrast) with 29.00% violation.

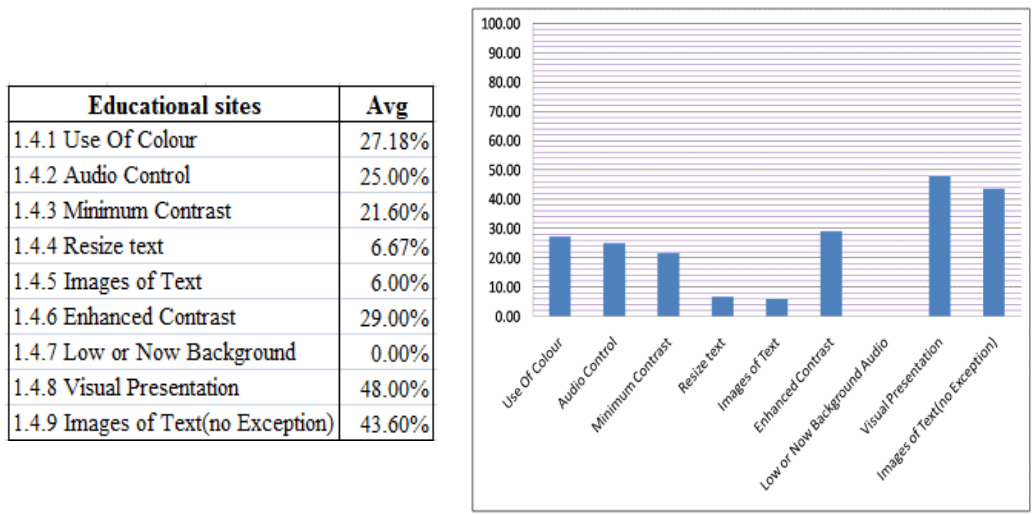

Figure 8: Violation of accessibility guideline (Distinguishable) in educational websites

International Journal of Computer Science, Engineering and Applications (IJCSEA) Vol.4, No.1, February 2014

**ii. Operable**

While considering operable principle in educational websites, the maximum violated guideline is guideline 2.1 (Keyboard accessible) in which educational websites indicate 55.00% violation. Educational websites show 6.25%, 35.00%, and 25.61% violation in guideline 2.2 (Enough Time), guideline 2.3 (Seizures), and guideline 2.4 (Navigable) respectively. Although the percentage is less but this rule is violated by approximately all websites, so we introduce a separate point by point graphical analysis in Figure 9. Here we find that the top 3 violated sub rules of navigability are guideline 2.4.8 (Location) with 47.00% violation, guideline 2.4.3 (Focus order) with 46.00% violation, and guideline 2.4.9 with 37.50% violation.

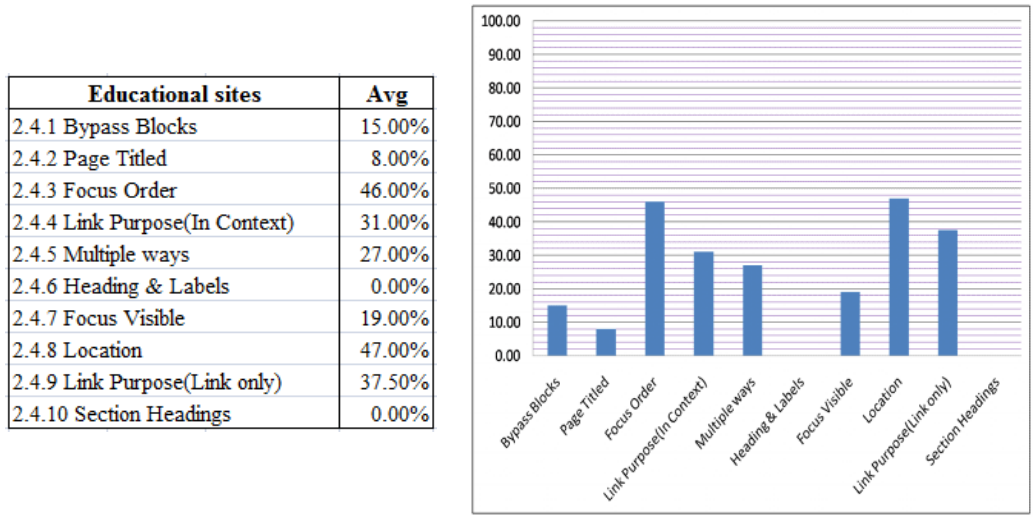

Figure 9: Violation of accessibility guideline (Navigable) in educational website

**iii. Understandable**

While considering understandable principle in educational websites, the maximum violated guideline is guideline 3.3 (Input Assistance) in which educational websites indicate 57.00% violation. Educational websites show 39.60% and 24.50% violation in guideline 3.1 (Readable) and Guideline 3.2 (Predictable) respectively. This rule is violated by approximately all websites, so we give a separate point by point graphical analysis in Figure 10. Here we find that the top 2 violated sub rules of predictability are guideline 3.2.5 (Change on request) with 50.00% violation and guideline 3.2.4 (Consistent Identification) with 37.50% violation.

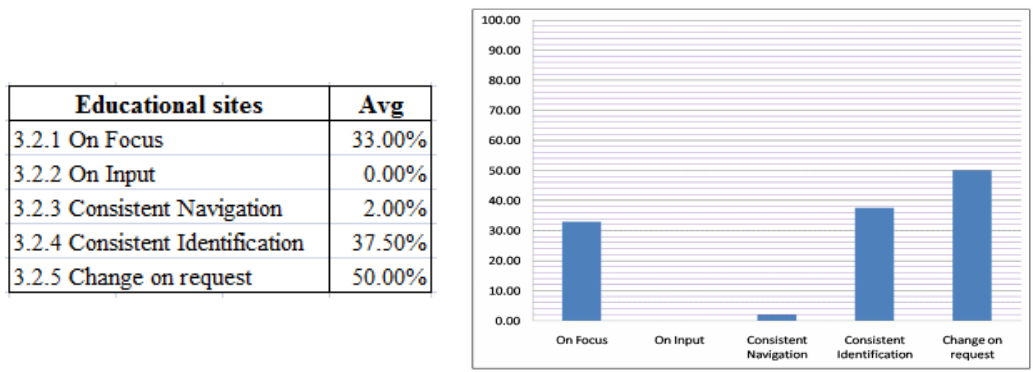

Figure 10: Violation of accessibility guideline (Predictable) in Educational websites

17

International Journal of Computer Science, Engineering and Applications (IJCSEA) Vol.4, No.1, February 2014

#### iv. Robust

According to our analysis the In Indian Educational Websites Guideline 4 (Robust) gives 10.00% violation.

### c. Commercial Sites

Figure 11.describes the overall average percentage violation results for each rule of WCAG 2.0. when applied to some of the Indian Commercial websites.

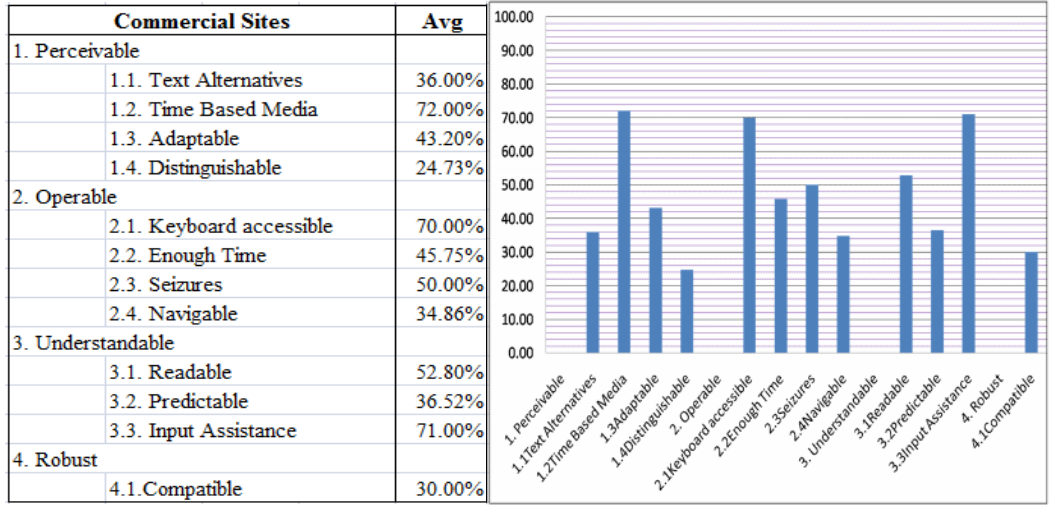

| Commercial Sites | Avg |
|---|---|
| 1. Perceivable | |
| 1.1. Text Alternatives | 36.00% |
| 1.2. Time Based Media | 72.00% |
| 1.3. Adaptable | 43.20% |
| 1.4. Distinguishable | 24.73% |
| 2. Operable | |
| 2.1. Keyboard accessible | 70.00% |
| 2.2. Enough Time | 45.75% |
| 2.3. Seizures | 50.00% |
| 2.4. Navigable | 34.86% |
| 3. Understandable | |
| 3.1. Readable | 52.80% |
| 3.2. Predictable | 36.52% |
| 3.3. Input Assistance | 71.00% |
| 4. Robust | |
| 4.1.Compatible | 30.00% |

Figure 11: Violation of accessibility guideline in Commercial websites

#### i. Perceivable

With respect to the perceivable principle in case of commercial websites, the maximum violated guideline is guideline 1.2 (Time Based Media) in which commercial websites show 72.00% violation. Commercial websites indicate 36.00%, 43.20%, and 24.73% violation in guideline 1.1 (Text Alternative), guideline 1.3 (Adaptable), and guideline 1.4 (Distinguishable) respectively. Although, the percentage is little less but this rule is violated by approximately all websites, so we introduce a separate point by point graphical analysis in Figure 12. Here we find that the top 3 violated sub rules for distinguishability are guideline 1.4.6 (Enhanced contrast) with 53.20% violation, guideline 1.4.3 (Minimum contrast) with 41.40% violation, and guideline 1.4.8 (Visual Presentation) with 36.00% violation.

International Journal of Computer Science, Engineering and Applications (IJCSEA) Vol.4, No.1, February 2014

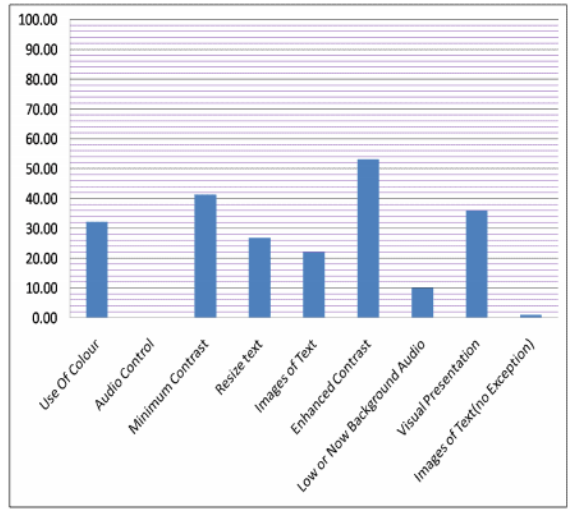

| Commercial Sites | Avg |
|---|---|
| 1.4.1 Use Of Colour | 32.15% |
| 1.4.2 Audio Control | 0.00% |
| 1.4.3 Minimum Contrast | 41.40% |
| 1.4.4 Resize text | 26.80% |
| 1.4.5 Images of Text | 22.00% |
| 1.4.6 Enhanced Contrast | 53.20% |
| 1.4.7 Low or Now Background | 10.00% |
| 1.4.8 Visual Presentation | 36.00% |
| 1.4.9 Images of Text(no Exception) | 1.00% |

Figure 12: Violation of accessibility guideline (Distinguishable) in commercial websites

### ii. Operable

In case of operable principle for commercial websites, the maximum violated guideline is guideline 2.1 (Keyboard accessible) in which commercial websites show 70.00% violation. Commercial websites indicate 45.75%, 50.00%, and 34.86% violation in guideline 2.2 (Enough Time), guideline 2.3 (Seizures), and guideline 2.4 (Navigable) respectively. Although, the percentage is less but this rule is violated by approximately all websites, so we introduce a separate point by point graphical analysis in Figure 13. Here we find that the top 3 violated sub rules of navigability are guideline 2.4.3 (Focus order) with 58.00% violation, guideline 2.4.9 with 56.75% violation, and guideline 2.4.7 (Focus Visible) with 41.00% violation.

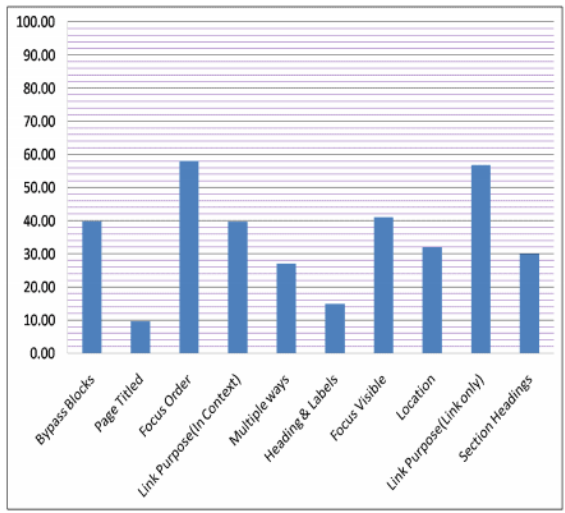

| Commercial Sites | Avg |
|---|---|
| 2.4.1 Bypass Blocks | 39.80% |
| 2.4.2 Page Titled | 9.60% |
| 2.4.3 Focus Order | 58.00% |
| 2.4.4 Link Purpose(In Context) | 39.60% |
| 2.4.5 Multiple ways | 27.00% |
| 2.4.6 Heading & Labels | 14.80% |
| 2.4.7 Focus Visible | 41.00% |
| 2.4.8 Location | 32.00% |
| 2.4.9 Link Purpose(Link only) | 56.75% |
| 2.4.10 Section Headings | 30.00% |

Figure 13: Violation of accessibility guideline (Navigable) in commercial websites

### iii. Understandable

While considering the understandable principle in commercial websites, the maximum violated guideline is guideline 3.3 (Input Assistance) in which commercial websites show 71.00% violation. Commercial websites indicate 52.80% and 36.52% violation in guideline 3.1 (Readable) and Guideline 3.2 (Predictable) respectively. This rule is violated by approximately all websites, so we introduce a separate point by point graphical analysis in Figure 14. Here we find that the top 2 violated sub rules of predictability are guideline 3.2.1 (On focus) with 58.00% violation and guideline 3.2.5 (Change on request) with 56.00% violation.

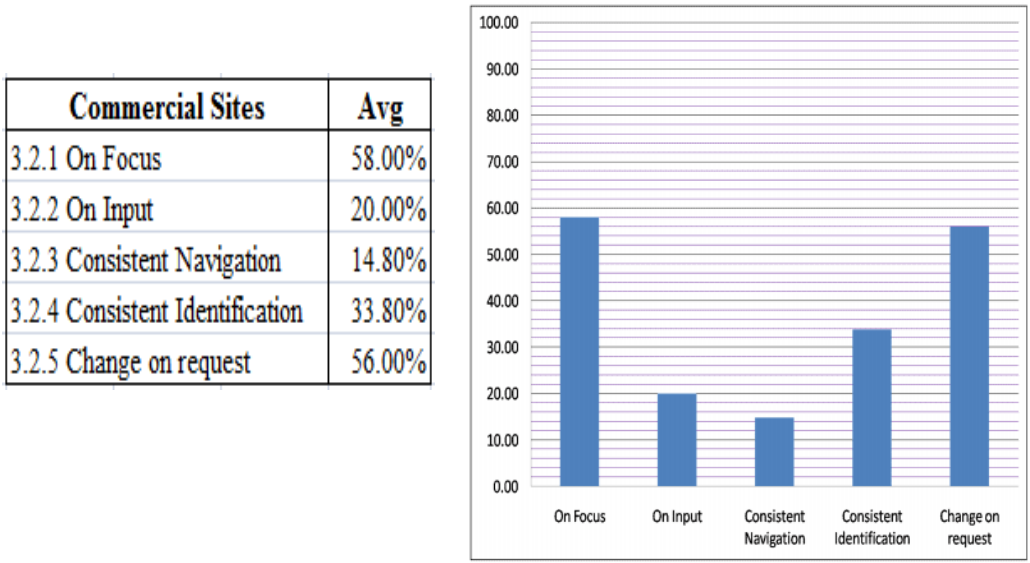

Figure 14: Violation of accessibility guideline (Predictable) in commercial websites

### iv. Robust

The Indian Commercial Websites considered in our study indicated 30.00% violation in case of Guideline 4 (Robust).

## 6. RESULT ANALYSIS

A comparative view of the violations with respect to the guidelines on Perceivable, Operable, Understandable, Robust for the three different categories of web portals considered in our study, namely, Government, Educational & commercial is presented in the pie charts of figure 15.

With respect to the Perceivable guideline most violation is found in Commercial websites (41.09%) and least violation is found in Educational Websites (26.91%). According to the rule of "text alternative" commercial, government as well as educational websites show 36.00%, 23.89% & 18.20% respectively (decreasing order of violation). But in case of "adaptable" rule educational websites (49.00%) and government websites (44.20%) indicate more violation than commercial websites (43.20%) as educational and government websites have lot of tables and in most cases table headers and table captions are missing. As per the rule of "visual presentation" government websites, educational websites and commercial websites show 60.00%, 48.00%, and 36.00% violation respectively. However, the sub rules under "visual presentation" are mostly violated by all web sites.

International Journal of Computer Science, Engineering and Applications (IJCSEA) Vol.4, No.1, February 2014

In case of the Operable guideline most violation is found in Commercial websites (44.93%) and least violation is found in Educational Websites (27.30%). According to the rule of keyboard accessibility commercial, educational, and government websites show 70.00%, 55.00%, and 37.00% respectively (in decreasing order of violation). As per the "location" rule educational websites (47.00%) and government websites (46.00%) show more violation than commercial websites (32.00%). The websites which have better compliance with respect to location are:

- India.gov.in (Among the government websites).
- www.uohyd.ac.in (Among the educational websites).
- www.flipkart.com (Among the commercial websites).

In case of "multiple ways" guidelines commercial websites, educational websites, and government websites show violation in decreasing order. According to this rule a website should contain any 2 of 5 points given below:

- A list of related pages
- Table of contents
- Site map
- Site search
- List of all available WebPages

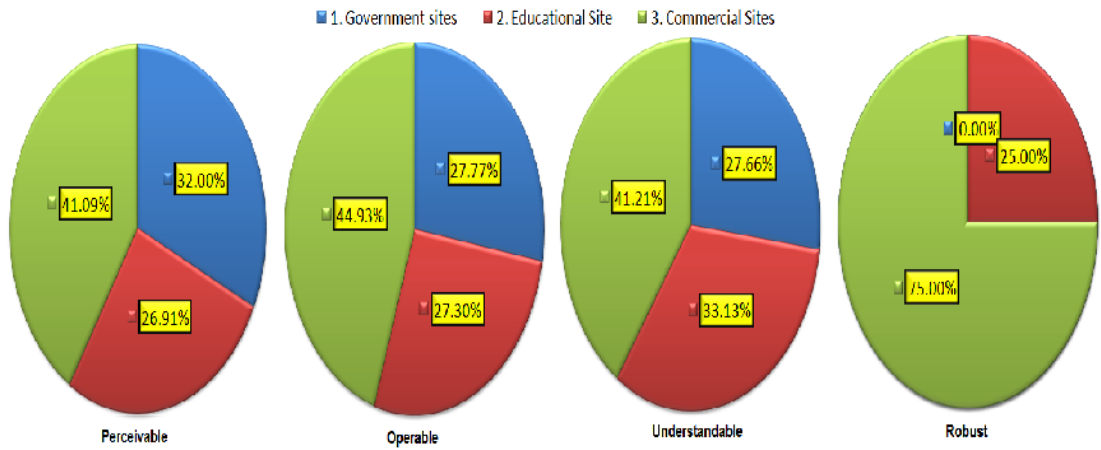

Figure 15: Violation of different websites

For the Understandability guideline most violation is found in Commercial websites (41.21%) and least violation is found in Government Websites (27.66%). According to the rule of "Input Assistance" commercial, government, and educational websites indicate 71.00%, 57.05% and 57.00% violation respectively (in decreasing order of violation). As per the Robustness guidelines most violation is found in Commercial websites (75.00%) and least violation is found in Government Websites (0%).

## 7. CONCLUSION

With the growing number of web portals worldwide, accessibility issues have emerged as a serious concern for web designers. In this paper, we have evaluated the accessibility of three major categories of web portals in the Indian context. The findings of our analysis clearly show the compliance of the web portals with respect to the WCAG 2.0 guidelines. We believe that the quantitative results of our evaluation can help web designers to incorporate the required features according to the WCAG 2.0 guidelines in order to make web portals more pragmatic and accessible to various user categories.

## AUTHORS

**Dr. Manas Ranjan Patra** holds a Ph.D. Degree in Computer Science and has been teaching for the last 25 years. Currently he is an Associate Professor and Director, Computer Centre at Berhampur University. He was a United Nations Fellow to IIST/UNU, Macao. He has more than 100 research publications to his credit. His research interests include Service Oriented Computing, Applications of Data mining and e-Governance. He has extensively travelled to many countries for presenting research papers, chairing technical sessions and delivering invited talks. He has been a member of Programme Committees and Editorial Boards of many International journals and conferences. 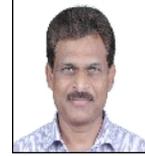

**Amar Ranjan Dash** holds a Bachelors degree in Technology and currently pursuing his Masters Programme in Computer Science at Berhampur University. His research interests include Web Technology, Cloud Computing and Fuzzy Logic. 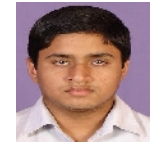

**Prasanna Kumar Mishra** holds a Bachelors degree in Technology. He worked as a Systems Analyst for three years before joining the Masters Programme in Computer Science. His research interests include Web Technology, Compiler Design and Cloud Computing. 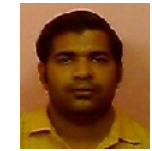

*INTENTIONAL BLANK*